\documentclass[12pt,floatfix,preprintnumbers]{revtex4}
\usepackage{graphicx}
\usepackage{dcolumn}
\usepackage{bm}
\usepackage{epsfig}
\usepackage{amsmath}
\usepackage{nicefrac} %
\usepackage{mathtools} %

\usepackage{mathrsfs} %

\begin{document}





%
\title{Laser Entropy} %
\author{Marlan O.~Scully} %
\affiliation{Texas A\&M University,~College~Station,~TX~77843}%
\affiliation{Princeton University, Princeton, NJ 08544}%
\affiliation{Baylor University, Waco, TX ,76798} %


\date{\today} %

\begin{abstract}
\begin{center} %
\textbf{Abstract} %
\end{center} %
The entropy of an ordinary (photon) laser and an atom laser (Bose condensate) is calculated. %
In particular, the nonzero entropy of a single mode laser or maser operating near threshold is obtained. %
This result is to be compared with the statement frequently made in the study of the maser heat engine %
to the effect that: %
``because maser radiation is in a pure state, its entropy is zero.'' %
Similarly, the entropy of the ground state of a Bose-Einstein condensate (a.k.a.~the atom laser) is also %
calculated for the first time. %
This is to be compared with the textbook wisdom which holds that: ``The condensed particles ... are condensed in momentum space, a set of stationary particles ... having zero energy and zero entropy.'' %
\end{abstract}

\maketitle

\newpage

\section{Introduction} %

Studying the entropy of thermal light led Planck 
to the quantum of action and Einstein to the photon concept. 
Half a century later the maser/laser appeared on the scene and it was shown that the three level maser could be regarded %
as a kind of quantum heat engine \cite{Ref:ScovilSchulzDeBois1959} yielding a quantum equivalent %
to the Carnot cycle \cite{Ref:Geusic1967}. %
More recently it has been recognized that quantum coherence in the lasing atoms allows lasing without inversion \cite{Ref:KocharovskayaPhyRep,Ref:HarrisPhysicsToday,Ref:ScullyZubairyQO}; %
and by extension that we can extract work from a \textit{single heat bath} (without violating the second law) via vanishing quantum coherence \cite{ref8}. %
A clear analysis of the maser as a quantum heat engine has been given \cite{Ref:TomasOpatrny}. %
As has an analysis of the Carnot bound on masers without inversion \cite{Ref:Kurizki} %
and laser cooling of solids \cite{Ref:Ruan}; %
for a recent review of quantum thermodynamics see \cite{Ref:Kosloff}.

\begin{figure} %
\includegraphics[width=0.8\textwidth]{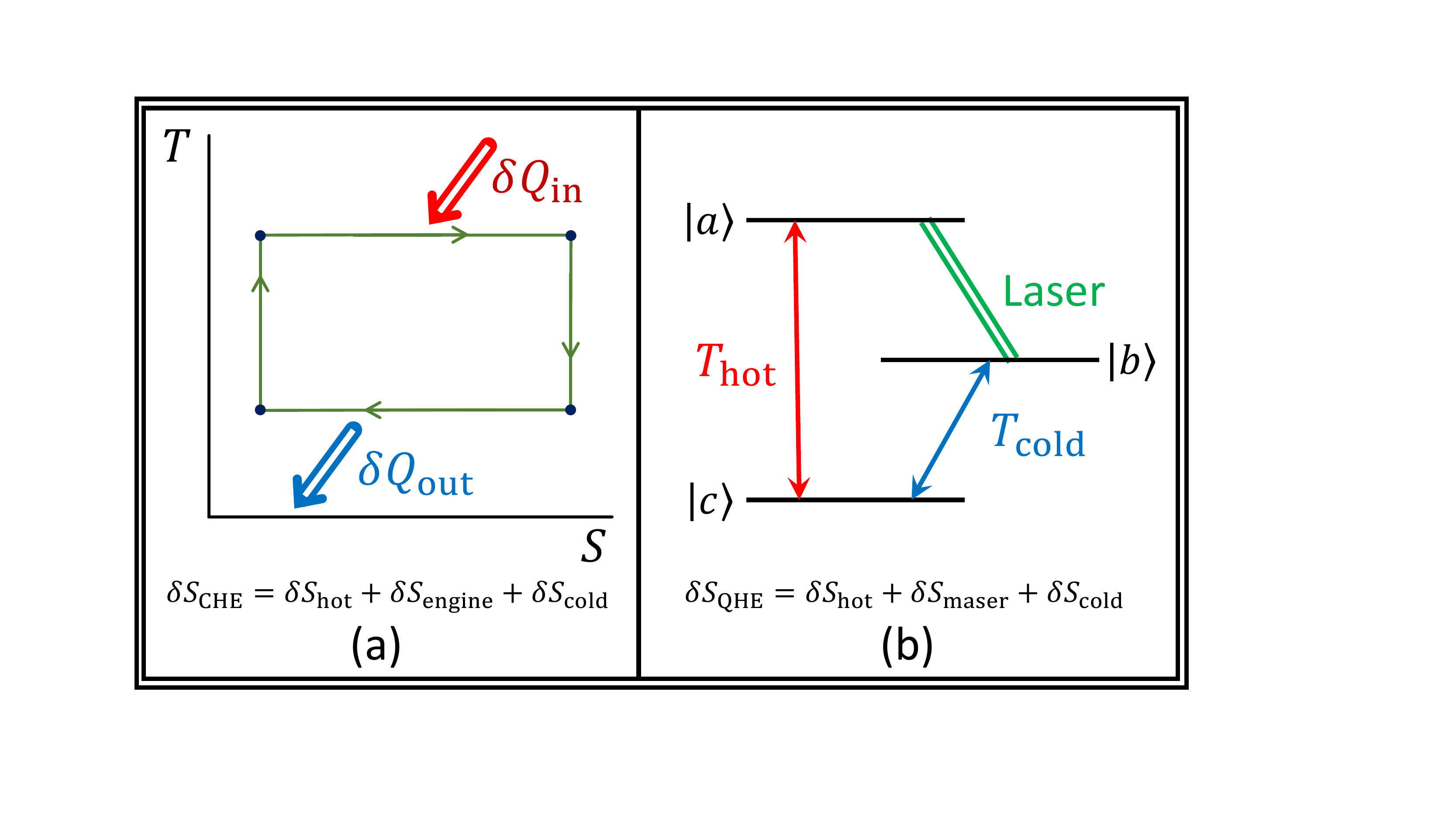}
\caption{(color online) (a) Classical Carnot heat engine (CHE) operates between high temperature energy source %
and low temperature energy sink. %
The entropy change for a complete cycle is the sum of contributions from the hot energy source and the cold entropy sink together with the entropy coming from the engine due to e.g.~friction. %
(b) The laser driven by hot and cold thermal reservoirs is a quantum heat engine (QHE). %
The entropy change for this QHE is the sum of single photon entropy changes due to the hot and cold light together with the contribution associated with the entropy change due to a single photon added to the laser/maser. %
} %
\label{Fig:Fig_Carnot} %
\end{figure}

The present work was initially stimulated by the studies of Harris \cite{Ref:HarrisPRA2016} %
on quantum heat engines and electromagnetically induced transparency \cite{Ref:HarrisPRL1990}, %
in which he shows that:
\begin{quote} %
``[U]sing the second law, one may easily obtain a result that using [the usual] Maxwell's and Schr\"{o}dinger's equations takes several pages of calculations.'' %
\end{quote} %
In particular, he uses an entropy relation similar to that in Ref.~\cite{Ref:ScovilSchulzDeBois1959} %
for a maser/laser system driven by hot and cold radiation, as in Fig.~\ref{Fig:Fig_Carnot}, %
given by \cite{Ref:EntropyChange} %
\begin{align} %
\label{Eq:EntropyChange} %
\delta S_\text{QHE} = -\frac{\hbar \nu_h}{T_h} + \delta S_\text{maser} + \frac{\hbar \nu_c}{T_c}, %
\end{align} %
where $\nu_h, T_h$ $[\nu_c, T_c]$ are the frequency and temperature of the hot [cold] monochromatic radiation resonant with the
$c \rightarrow a$ $[c \rightarrow b]$ transition, %
and $\delta S_\text{maser}$ is the maser entropy change associated with a change in the average photon number of one. %

The physics behind Eq.~\eqref{Eq:EntropyChange} is similar to the textbook treatment %
of the classical Carnot heat engine (CHE) of Fig.~\ref{Fig:Fig_Carnot}; %
in which the entropy change after a complete cycle
$\delta S_\text{CHE}$ as determined by drawing energy $\delta Q_\text{in}$ %
from a high temperature energy source and dumping energy $\delta Q_\text{out}$ %
into a low tempeature entropy sink is given by %
\begin{align} %
\label{Eq:EntropyCHE} %
\delta S_\text{CHE} = \frac{\delta Q_\text{in}}{T_h} + \delta S_\text{engine} + \frac{\delta Q_\text{out}}{T_c}, %
\end{align} %
where $\delta S_\text{engine}$ is the entropy generated by engine inefficiency, e.g., friction. %
By conservation of energy %
the work $\delta W = \delta Q_\text{in} - \delta Q_\text{out}$ %
and so we have the famous Carnot efficiency %
\begin{align} %
\frac{\delta W}{\delta Q_\text{in}} \leq 1 - \frac{T_c}{T_h}. %
\end{align} %

Equation \eqref{Eq:EntropyChange} is similar in spirit to Eq.~\eqref{Eq:EntropyCHE} %
and as was shown in \cite{Ref:Geusic1967}, %
the change in entropy corresponding to a single hot (energy source) photon absorbed and a maser photon emitted together with a cold (entropy sink) photon is given by Eq.~\eqref{Eq:EntropyChange}. %
Now at threshold, the populations in $|a\rangle$ and $|b\rangle$ are equal %
so the entropy change per photon of the maser $\delta S_m = \hbar \nu_m / T_m$ is said %
\cite{Ref:ScovilSchulzDeBois1959, Ref:Geusic1967} to vanish since $T_m = \infty$. %
In general Eq.~\eqref{Eq:EntropyChange} yields the Carnot quantum efficiency %
\begin{align} %
\label{Eq:CarnotQuantumEfficiency} %
\frac{\hbar \nu_m}{\hbar \nu_h}\leq 1 -\frac{T_c}{T_h},
\end{align} %
where we have used the fact that $\hbar \nu_m = \hbar \nu_h - \hbar \nu_c$ %
This result is a good example in support of Harris' point since the derivation %
of Eq.~\eqref{Eq:CarnotQuantumEfficiency} %
by conventional density matrix techniques \cite{Ref:ScullyPNAS} takes a bit of algebra. %

Equation \eqref{Eq:EntropyChange} clearly applies below threshold %
when the emitted ``laser'' light is essentially thermal. %
But what if we are above threshold? %
One often encounters statements such as: %
``because the maser radiation is in a pure state, its entropy is zero.'' %
But the maser/laser radiation is not in a pure state. %
And, as is shown in section \ref{Sec:LaserEntropy} and the appendix, the entropy of maser light is not zero %
but is determined by the density matrix formulation of the quantum theory of the laser \cite{Ref:QTL} %
in which the photon and atom laser statistics is calculated is based on the master equation %
\begin{align} %
\label{Eq:MasterEquationAtomLaser} %
\dot{\rho}_{n,n} = - G(n+1) \, (n+1) \rho_{n,n} + G(n) \, n \rho_{n-1,n-1} %
- L n \rho_{n,n} + L(n+1) \rho_{n+1,n+1}, %
\end{align} %
where the gain and loss coefficients, $G$ and $L$, for the photon and atom lasers are given in sections \ref{Sec:LaserEntropy} and \ref{Sec:BECEntropy}. %

In the next section we sketch the calculation of the laser/maser entropy %
from the density matrix formulation of the quantum theory of the optical maser \cite{Ref:QTL} %
and compare it to the entropy of high temperature single mode thermal light, i.e.~the maser below threshold. %
The maser entropy flux is calculated and compared to that of thermal light as well as that of a damped coherent state. %
In section \ref{Sec:DiscussionSummary}, %
the entropy of the analogous ground state of a Bose Einstein \cite{Ref:Meystre} condensate, %
a.k.a.~the atom laser is presented. %
A summary and discussion is given in Section \ref{Sec:DiscussionSummary}. %

\section{Laser entropy} %
\label{Sec:LaserEntropy} %

In the quantum theory of the optical maser the gain coefficient $G(n)$ and the loss rate $L$ of Eq.~\eqref{Eq:MasterEquationAtomLaser} %
are given by %
\begin{align} %
\label{Eq:LaserGain} %
G(n) = \frac{\alpha}{1 + \frac{\beta}{\alpha}n} \quad \text{and} \quad L = \gamma, %
\end{align} %
in terms of the laser parameter: $\alpha = $ linear gain, $\beta = $ nonlinear saturation coefficient, %
and the cavity loss rate $\gamma = \nu / Q$ is governed by the cavity $Q$ factor. %
As is shown in \cite{Ref:QTL}, the steady state solution to \eqref{Eq:MasterEquationAtomLaser} yields the $n$ photon probability distribution which can be written as %
\begin{subequations} %
\begin{align} %
\rho_{nn} = \frac{1}{Z} \frac{B! A^n}{(n+B)!}, %
\end{align} %
where the normalization is given in terms of the confluent hypergeometric function %
\begin{align} %
Z = \prescript{}{1}{F}_1 (1; B+1; A), %
\end{align} %
and
\begin{align} %
A = \frac{\alpha^2}{\beta\gamma} \quad \text{and} \quad B = \frac{\alpha}{\beta}. %
\end{align} %
\end{subequations} %

Far enough above threshold, e.g. $(\alpha - \gamma)/\gamma \gtrsim 0.1$, %
we may write $\rho_{nn}$ in the appealing form %
\begin{align} %
\label{Eq:LaserDM_Photon} %
\rho_{nn} = \frac{A^{n+B}}{(n+B)!} e^{-A}. %
\end{align} %

We also note that far enough above threshold the photon distribution can be approximated by the Gaussian %
\begin{align} %
\label{Eq:LaserDMGaussian} %
\rho_{n,n} \cong \frac{1}{\sqrt{2\pi A}} \exp \left[ - \frac{(n-\overline{n})^2}{2 A} \right], %
\end{align} %
where $\overline{n} = A - B = \tfrac{\alpha}{\gamma} \tfrac{\alpha - \gamma}{\gamma}$. %

Plugging $\rho_{nn}$ given by Eq.~\eqref{Eq:LaserDM_Photon} and/or \eqref{Eq:LaserDMGaussian} %
into the von Neumann entropy equation %
\begin{align} %
\label{Eq:vonNeumannEntropy} %
S = - k_\text{B} \sum_n \rho_{nn} \ln \rho_{nn}, %
\end{align} %
we obtain %
\begin{align} %
\label{Eq:LaserEntropyAboveT} %
S \cong k_\text{B} \ln \sqrt{ 2\pi \frac{\alpha}{\alpha - \gamma} \overline{n}} + \frac{k_\text{B}}{2}. %
\end{align} %
We note that the entropy flux implied by Eq.~\eqref{Eq:LaserEntropyAboveT} is given by %
\begin{align} %
\label{Eq:MaserEntropyFlux} %
\dot{S}_\text{maser} = \frac{k_\text{B} \dot{\overline{n}}}{ 2 \overline{n}_m} = \frac{\kappa}{2 \overline{n}_m}, %
\end{align} %
where $\kappa = k_\text{B} P/\hbar \nu_\ell$ and $P$ is the emitted power. %
The preceding is to be compared with monochromatic thermal light characterized by the density matrix
\begin{align} %
\label{Eq:DMThermalLight} %
\rho_{nn} = \frac{ \overline{n}^{n}}{(\overline{n}+1)^{(n+1)}},
\end{align} %
where $\overline{n}$ is the Planck function
\begin{align} %
\overline{n}=\frac{1}{\exp(\hbar\nu/k_\text{B} T)-1}.
\label{Eqn11}
\end{align} %
In this case, Eq.'s \eqref{Eq:vonNeumannEntropy} and \eqref{Eq:DMThermalLight} yield the black-body entropy
\begin{align} %
\label{Eq:EntropyThermal} %
S_\text{thermal} = k_\text{B} (\overline{n}+1)\ln(\overline{n}+1) - k_\text{B} \overline{n}\ln \overline{n}, %
\end{align} %
which in the high temperature limit is %
\begin{align} %
S = k_\text{B} \ln \overline{n}_\text{high} + k_\text{B}, %
\end{align} %
where $\overline{n}_\text{high} = k_\text{B} T / \hbar \nu$, and the associated entropy flux is %
\begin{align} %
\label{Eq:EntropyFluxThermal} %
\dot{S}_{\text{thermal}} = \frac{\hbar \nu}{T}\dot{\overline{n}} %
= \frac{\kappa}{ \overline{n}_\text{high}}. %
\end{align} %
Finally, we note that for a laser well below threshold $G = \alpha$ and $S$ is given by Eq.~\eqref{Eq:EntropyThermal} %
with $\overline{n} = [ (\alpha/ \gamma) - 1]^{-1}$. %

\section{BEC (\MakeLowercase{a.k.a.~}``Atom Laser'') Entropy} %
\label{Sec:BECEntropy} %

Bose Einstein Condensation (BEC) has been dubbed the ``atom laser'' \cite{Ref:KleppnerPhysicsToday} %
and it has been shown that the density matrix treatment for the photons in a laser cavity given by Eq.~\eqref{Eq:MasterEquationAtomLaser} also applies to the ground state of the BEC. %
In this case the index $n$ is replaced by $n_0$ denoting the number of atoms in the lowest state %
having energy $\epsilon_0$. %
Einstein taught us that for $N$ atoms in a box the average number in the condensate is %
$\overline{n_0} = N ( 1 - (T/T_c)^3)$ where $T_c$ is the critical temperature \cite{Ref:AvgNexponent}. %

We are here interested in the probability of having $n_0$ out of $N$ in the ground state of a parabolic trap
for which $\overline{n_0} = N ( 1 - (T/T_c)^3)$. %
This probability is given by the diagonal elements of the ground state density matrix $\rho_{n_0, n_0}$ %
which obeys Eq.~\eqref{Eq:MasterEquationAtomLaser} with gain %
\begin{align} %
\label{Eq:BECGain} %
G(n_0) = \kappa (N - n_0) %
\end{align} %
describing the rate of addition of atoms (gain) to the ground state due to the excited atoms %
($\epsilon_k, k \neq 0$) colliding with the walls having temperature $T$ and falling into the ground state at a rate $\kappa$. %
Likewise atoms are removed (lost) from the ground state due to interaction with the hot walls (temperature $T$) at a rate %
\begin{align} %
\label{Eq:BECLoss} %
L(n_0) = \kappa N (T / T_c)^3. %
\end{align} %
The master equation for $\rho_{n_0,n_0}$ obtained from Eq.'s~(\ref{Eq:MasterEquationAtomLaser}, \ref{Eq:BECGain}, \ref{Eq:BECLoss}) has the steady state ($\dot{\rho}_{n_0,n_0} = 0$) solution given by \cite{Ref:ScullyCNB1} %
\begin{align} %
\label{Eq:DMGndBEC} %
\rho_{n_0,n_0} = \frac{\mathcal{H}^{N-n_0}}{(N-n_0)!} e^{-\mathcal{H}}, %
\end{align} %
where $\mathcal{H} = N (T/T_c)^3$. %

The BEC ground state entropy obtained by inserting \eqref{Eq:DMGndBEC} into Eq.~\eqref{Eq:vonNeumannEntropy} %
can be plotted as a function of $T/T_c$; %
the result is found to be in good agreement with the ground state entropy obtained from exact numerical calculations for a mesoscopic condensate of say $10^3$ atoms. %

Here we will simply note that for low-enough temperatures the variance of the BEC atom distribution Eq.~\eqref{Eq:DMGndBEC} %
is governed to a reasonable approximation by $N (T/T_C)^3$ \cite{Ref:ScullyCNB1} %
and the BEC ground state entropy for a parabolic trap is found to be \cite{Ref:EntropyZeroTemp} %
\begin{align} %
\label{Eq:BECGroundStateEntropy} %
S_g = k_\text{B} \ln \sqrt{ 2 \pi N (T/T_C)^3} + \frac{k_\text{B}}{2}. %
\end{align} %

\section{Discussion and Summary} %
\label{Sec:DiscussionSummary} %

We now turn to a discussion and summary of our results. %

1.~Back to the quantum heat engine: %
To begin with, let us return to Eq.~\eqref{Eq:EntropyChange} %
and the question: ``What should we take for $\delta S_m$ in Eq.~\eqref{Eq:EntropyChange}?''
We note that the change in entropy due to a single photon addition or subtraction is obtained from Eq.'s \eqref{Eq:MaserEntropyFlux} and \eqref{Eq:EntropyFluxThermal} using the replacement %
$\delta n = \dot{\overline{n}} \delta t = \pm 1$ to obtain %
\begin{align} %
\nonumber %
\delta S_\text{maser} = \frac{k_B}{2\overline{n}_m} \quad \text{and} \quad
\delta S_\text{thermal} = \frac{k_B}{\overline{n}_\text{high}}, %
\end{align} %
where $\overline{n}_\text{high} = k_B T / \hbar \nu$ is the number of thermal photons in the high temperature limit. %

Hence, for a laser very near threshold with $\overline{n} \sim 10^6$ say, %
then $\delta S_\text{laser} \sim 10^{-6} k_B$. %
If this is compared with the thermal entropy change expression $k_B / \overline{n}_\text{high}$, %
and $\overline{n}_\text{high} = k_B T / \hbar \nu \sim 1$ for $k_BT$ and $\hbar \nu$ %
both around $1 \, \text{eV}$, we see that in this case $\delta S_\text{laser}$ is negligible, %
and the Carnot efficiency result of Eq.~\eqref{Eq:CarnotQuantumEfficiency} is valid. %

However, for a maser with $\hbar \nu \sim 10^{-6}\, \text{eV}$ and $k_B T \sim 1 \, \text{eV}$, $\overline{n}_\text{high} \sim 10^6$. %
So if $\overline{n}_\text{maser} \sim 10^6$ then $\delta S_\text{maser}$ is comparable to the entropy change $\delta S_\text{thermal}$. %
In such a case $\delta S_\text{maser}$ is not negligible. %
Eq.~\eqref{Eq:CarnotQuantumEfficiency} assumes $\delta S_\text{maser} \cong 0$ compared to $\delta S_\text{thermal}$ but this need not always be the case. %
Clearly the laser/maser entropy change per cycle depends on the specific scenario. %
In general the entropy flux equation for our maser atom problem driven by hot and cold radiation, %
as in Fig.~\ref{Fig:Fig_Carnot}, is %
\begin{subequations}
\begin{align} %
\dot{S}_\text{h} + \dot{S}_\text{m} + \dot{S}_\text{c} \geq 0, %
\end{align} %
which in view of Eq.'s (\ref{Eq:MaserEntropyFlux}, \ref{Eq:EntropyFluxThermal}) reads %
\begin{align} %
\label{Eq:MaserEntropyIncrese} %
\frac{\hbar \nu_h}{T_h} \dot{\bar{n}}_h +
\frac{k_\text{B}}{2\bar{n}_m} \dot{\bar{n}}_m +
\frac{\hbar \nu_c}{T_c} \dot{\bar{n}}_c \geq 0, %
\end{align} %
\end{subequations}
and since $-\dot{n}_h = \dot{n}_m = \dot{n}_c$, using $\nu_c = \nu_h - \nu_m$ %
and for high temperatures $\hbar \nu_h / T_h = k_\text{B}/n_h$, %
the maser entropy term in \eqref{Eq:MaserEntropyIncrese} is negligible since $\overline{n}_m \gg \overline{n}_h$. %
Thus we are again led to Eq.~\eqref{Eq:CarnotQuantumEfficiency} %
even though we are now above threshold. %
But for some problems such as the micromaser \cite{Ref:Haroche} %
$\overline{n}_m$ is not a large number. %
We leave this as an open problem to be treated elsewhere. %

%
2.~The entropy is not given by a simple $S = k_B \ln W$ type expression given by Eq.~\eqref{Eq:LaserEntropyAboveT} %
in the threshold region. %
Well above threshold i.e.~for $(\alpha - \gamma)/\gamma \gtrsim 0.1$ the simple form given %
by Eq.~\eqref{Eq:DMGndBEC} does obtain as does the entropy flux given by Eq.~\eqref{Eq:MaserEntropyFlux}. %

\begin{figure} %
\centering %
\includegraphics[width=0.85\textwidth]{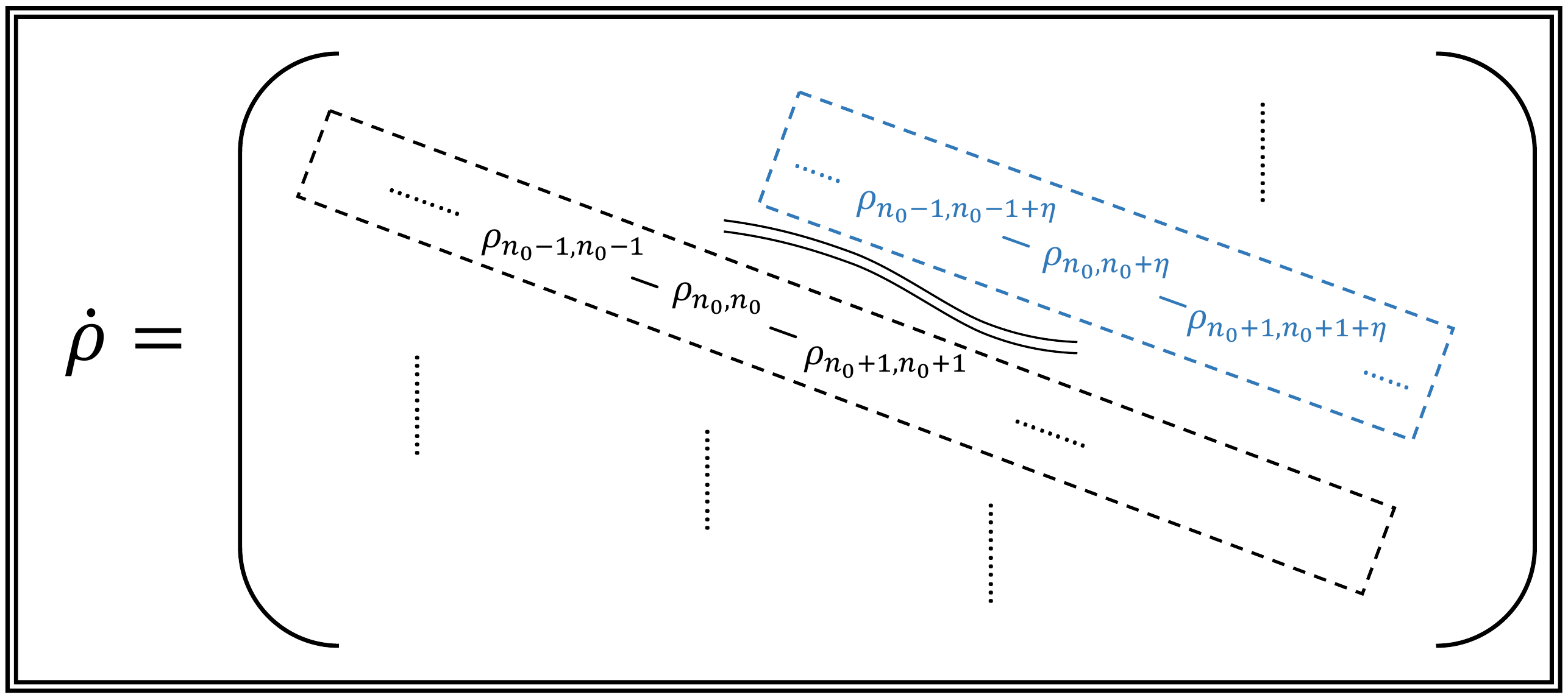} %
\caption{(color online) The density matrix equation of motion couples only elements of equal off diagonality $\eta$. %
For example in Eq.~\eqref{Eq:MasterEquationAtomLaser} for the photon statistics $\eta = 0$. %
} %
\label{Fig:Offdiagonal_DM} %
\end{figure} %

3.~Laser Entropy and the laser linewidth: %
Indeed, the source of the maser entropy is presaged by the insightful statement of Morse \cite{Ref:Morse} who says: %
``during a spontaneous process $\cdots$ the entropy always increases.'' %
In fact it is precisely the spontaneous (as opposed to stimulated) emission events %
which are the source of the Schawlow-Townes optical maser linewidth; %
and which are the source of the time dependence of the laser radiation density matrix given by %
\cite{Ref:ScullyCNB1}
\begin{align} %
\label{Eq:TimeDependenceDO} %
\rho_{n,n+\eta}(t) = \rho_{n,n+\eta} (0) e^{-\eta^2 D t} %
\end{align} %
where $\eta$ measures the degree of off-diagonality as per Fig.~\ref{Fig:Offdiagonal_DM}. %
Eq.~\eqref{Eq:TimeDependenceDO} implies the electric field %
\begin{align} %
\label{Eq:TimeDependenceElectricField} %
\langle \hat{E} (t) \rangle = \sum_n \, \mathscr{E}_0 \, \rho_{n,n+1} (0) \, \sqrt{n+1} \, \exp \left( i \nu t - D t \right), %
\end{align} %
where $\mathscr{E}_0$ is the electric field per photon and $D \equiv \alpha / 4 \overline{n}$ is the laser phase diffusion coefficient \cite{Ref:Stanley}. %
The Fourier transform of Eq.~\eqref{Eq:TimeDependenceElectricField} is a Lorentzian centered at $\nu_l$ %
and with a full width at half max given by %
\begin{align} %
\label{Eq:LaserLinewidth} %
\Delta \nu = 2 D = \frac{\alpha}{2\overline{n}}. %
\end{align} %

The physics behind the linewidth \eqref{Eq:LaserLinewidth} is (partially) illustrated %
by writing the equations of motion for a maser below threshold as %
\begin{subequations} %
\begin{align} %
\label{Eq:EqnMotionPhotonNumberBelowThreshold} %
\dot{\overline{n}} = & \, \alpha ( \overline{n} + 1) - \gamma \overline{n}, \\ %
\label{Eq:EqnMotionElectricFieldBelowThreshold} %
\dot{\overline{E}} = & \, \frac{1}{2} ( \alpha - \gamma) \overline{E}, %
\end{align} %
\end{subequations} %
where we here use the notation $\overline{E} = \langle E \rangle$. %
Then, below threshold, the steady state relation \eqref{Eq:EqnMotionPhotonNumberBelowThreshold} yields %
$\alpha - \gamma = \alpha / \overline{n}$ %
and using this in \eqref{Eq:EqnMotionElectricFieldBelowThreshold} yields %
$\dot{\overline{E}} = - (\alpha / 2 \overline{n}) \overline{E}$; %
which implies a phase diffusion coefficient $D' = \alpha / 2 \overline{n}$ and a below threshold linewidth %
\begin{align} %
\Delta \nu' = 2 D' = \frac{\alpha}{\overline{n}}. %
\end{align} %
Concluding this linewidth review %
we note that $\alpha$ is essentially $\gamma = \nu / Q$ in steady-state; %
and we compare the proceeding linewidth discussion with the ``spontaneously generated'' laser entropy flux
below and above threshold in table \ref{Table:LaserLinewidthEntropy}. %

\begin{table} %
\begin{tabular}{c|c|c|} %
& below theshold & above threshold \\ %
\hline %
laser linewidth & $\frac{\nu / Q}{\overline{n}_h}$ & $\frac{\nu / Q}{2 \overline{n}_l}$ \\ %
\hline %
entropy flux & $\frac{\kappa}{\overline{n}_h}$ & $\frac{\kappa}{2 \overline{n}_l}$ \\ %
\hline %
\end{tabular} %
\caption{The spontaneously generated entropy flux for hot thermal light ($k_\text{B}T \gg \hbar \nu$) %
having average photon number $\overline{n}_h = k_\text{B}T / \hbar \nu$ with the flux of a laser above threshold %
having average photon number $\overline{n}_l$ where $\kappa$ is defined following Eq.~\eqref{Eq:MaserEntropyFlux}. %
This is compared with the laser linewidth below and above threshold %
which are well known to differ by a historically bothersome factor of two, the origin of which is clear in the maser entropy flux.} %
\label{Table:LaserLinewidthEntropy} %
\end{table} %

4.~Off-diagonality and more: %
Several points should be made concerning the off-diagonal nature of the maser density matrix, a few of these are: %

\begin{enumerate} %
\item[(\textit{i)}] As is seen from Eq.~\eqref{Eq:LaserEntropyAboveT}, %
the maser entropy well above threshold takes the form of the famous Boltzmann microcanonical entropy but is quite different; %
for example, the entropy of a gas is extensive (i.e.~goes as the number of gas atoms) but the maser entropy is not an extensive variable. %
\item[(\textit{ii)}] The factor of $2$ in the linewidth encountered in going from below to above threshold is %
a well known, if a bit subtle, aspect of laser physics. %
On the other hand the laser entropy flux, factor of $2$ in passing through threshold is due to the laser entropy going from $\ln \overline{n}$ to $\ln \sqrt{ \overline{n}}$. %
\item[(\textit{iii)}] The degree of off-diagonality $\eta$ as it appears in Eq.~\eqref{Eq:TimeDependenceDO} can be large %
i.e.~$0 \leq \eta \leq n$ where $n$ can be of order ( or greater than) $\overline{n}$. %
Hence such off-diagonal character of the laser density matrix vanishes rapidly %
as is shown by Eq.~\eqref{Eq:TimeDependenceDO}.
The paper by Chen and Fan \cite{Ref:ChenFan2013} treats the off-diagonal term %
but uses a linear gain-loss master equation. %
\end{enumerate} %

5.~On the entropy of the BEC ground state entropy: %
Finally we note that in the thermodynamic limit the entropy of a Bose gas \cite{Ref:RomeroRochin} has the dependence %
\begin{align} %
S \approx 3.6 k_B N \left( \frac{T}{T_C} \right)^3. %
\end{align} %
For a macroscopic Bose gas of say $10^{23}$ atoms $S_g \sim \ln N$ is negligible compared to $S$. %
But for a mesoscopic BEC of $10^3$ atoms $N (T/T_c)^3 = 1$ when $T/T_c \cong 0.1$; %
and in such a case $S \sim 4 k_\text{B}$ and $S_g \sim k_\text{B}$ are of the same order.

We emphasize that the present BEC entropy analysis is approximate but as will be further discussed elsewhere, %
it gives a good account of the ground state entropy. %
This is to be compared with conventional wisdom which one often hears saying that \cite{Ref:Morse}: %
\begin{quote} %
As expected, the $n_0$ particles constituting the ``condensate'' do not contribute to the entropy of the system, %
while the $N - n_0$ particles that constitute the normal part do contribute. %
\end{quote} %
The ground state entropy of a mesoscopic BEC yields many interesting questions. %
For example, the relation between the correlation entropy and the ground state entropy is an open question. %

6.~Summary: The quantum entropy of a laser below, at, and above threshold is well described by the quantum theory of the maser. %
The entropy flux of the maser is not ``zero'' and this can be important for a complete analysis of the Carnot bound of maser operation. %
A similar analysis of the quantum theory of the ``atom laser'' yields a nonvanishing BEC ground state entropy. %

The present paper poses many open questions, for example: %

\begin{enumerate} %
\item[(\textit{i})] The threshold $\alpha = \gamma$ region is interesting and should be further investigated. %
\item[(\textit{ii})] It would be interesting to extend the laser entropy -- linewidth discussion %
to include the noise generated correlated emission laser and lasing in the presence of squeezed light. %
\item[(\textit{iii})] The treatment of the full nonlinear master equation \cite{Ref:ScullyCNB1} is challenging. %
\item[(\textit{iv})] The ground state entropy is \textit{not} simply the total entropy minus %
the excited state entropy, as will be shown elsewhere. %
\end{enumerate} %

\begin{acknowledgments} %
The support of the ONR (Award No.~N00014-16-1-3054) and the Welch Foundation (Grant No.~A-1261) has been instrumental in enabling this work. %
The author also thanks S.~Harris for stimulating this research %
and G.~Agarwal, M.~Kim,  V.~Kocharovsky, M. Shlesinger, and A.~Svidzinsky for helpful discussions. %
Work with J.~Ben-Benjamin, H.~Dong, S.~Li, R.~Nessler, and H.~Eleuch will be reported elsewhere. %
Special thanks go to M.~Kim for his excellent help in typing the manuscript and preparing the figures. %
\end{acknowledgments} %

\section*{Appendix: Laser Entropy Details} %

\renewcommand{\theequation}{A.\arabic{equation}} %
\setcounter{equation}{0} %

From the quantum theory of the laser \cite{Ref:ScullyZubairyQO,Ref:QTL} we have %
\begin{align} %
\label{Eq:QTLaserDM} %
\rho_{nn} = \frac{1}{Z} \frac{ \left( \frac{\alpha}{\beta} \right) ! \left( \frac{\alpha^2}{\beta \gamma} \right)^n}{ \left( n + \frac{\alpha}{\beta} \right)!} %
\end{align} %
where the normalization is expressed in terms of the confluent hypergeometric function as %
\begin{align} %
Z = \prescript{}{1}{F}_1 \left( 1; \frac{\alpha}{\beta} + 1; \frac{\alpha^2}{\beta\gamma} \right). %
\end{align} %
The $\alpha$, $\beta$, $\gamma$ laser parameters are defined in the text following Eq.~(6). 
In the usual laser limit of large $\alpha^2/\beta\gamma$ we have %
\begin{align} %
\prescript{}{1}{F}_1 \left( 1; \frac{\alpha}{\beta} + 1; \frac{\alpha^2}{\beta\gamma} \right) \Rightarrow \left( \frac{\alpha}{\beta} \right)! e^{\alpha^2/\beta\gamma} %
\left( \frac{\alpha^2}{\beta\gamma} \right)^{-\tfrac{\alpha}{\beta}}, %
\end{align} %
which inserted into \eqref{Eq:QTLaserDM} yields Eq.~(8). 

To calculate the entropy we write the entropy %
\begin{align} %
\nonumber %
S = & \, - k_\text{B} \sum_n \rho_{nn} \ln \rho_{nn}, \\ %
\intertext{using Eq.~(8) as} %
\label{Eq:ExpandLn} %
= & \, - k_\text{B} \sum_n \rho_{nn} \left[ (n + B) \ln A - A - \ln (n + B)! \right]. \\ %
\intertext{Making use of Stirling's approximation and noting that $A = B + \overline{n}$ we have} %
= & \, k_\text{B} \sum_n \rho_{nn} \left[ \ln \sqrt{ 2\pi (n+B)} + (n+B) \big( \ln (n+B) - \ln A \big) \right], %
\end{align} %
and expanding the logarithms to second order %
\begin{align} %
\label{Eq:ExpandApproxB} %
\ln (n+B) = \ln (A + n - \overline{n}) \cong \ln A + \frac{n-\overline{n}}{A} - \frac{(n-\overline{n})^2}{2A^2}, %
\end{align} %
we arrive at %
\begin{align} %
S \cong k_\text{B} \ln \sqrt{ 2\pi A} + \frac{k_\text{B}}{2}. %
\end{align} %

Similarly, as will be discussed in detail elsewhere, the BEC ground state density matrix can be written as %
\begin{align} %
\label{Eq:DMGndBECAppend} %
\rho_{n_0,n_0} = \frac{ \mathscr{H}^{N-n_0}}{ (N - n_0)!} e^{-\mathscr{H}}, %
\end{align} %
which leads to a ground state entropy %
\begin{align} %
\label{Eq:EntropyGndBECAppend} %
S_g = k_\text{B} \ln W_g + \frac{k_\text{B}}{2}, \qquad W_g = \sqrt{ 2\pi \mathscr{H}}, %
\end{align} %
where $\mathscr{H} = N ( T / T_c)^3$. %
Equation \eqref{Eq:EntropyGndBECAppend} is correct over a wide range of temperature, %
however Eq.~\eqref{Eq:DMGndBECAppend} shows that $\rho_{n_0,n_0} \sim \delta_{N,n_0}$ and the entropy vanishes at $T = 0$. %
It should be noted that the BEC ground state entropy \eqref{Eq:EntropyGndBECAppend} is not the total entropy minus the excited state entropy as will be discussed at length elsewhere. %


\begin{thebibliography}{99}
\bibitem{Ref:EntropyChange} %
Eq.~\eqref{Eq:EntropyChange} describes the entropy changes when a single hot (pump) photon is absorbed and a maser and cold (entropy sink) photon are emitted. %
\bibitem{Ref:AvgNexponent} %
Einstein considered the case of a square well trap for which the $(T/T_c)^3$ factor becomes $(T/T_c)^{3/2}$. %
\bibitem{Ref:EntropyZeroTemp} %
Eq.~\eqref{Eq:BECGroundStateEntropy} applies when we are not too close to $T = 0$. %
When we are at $T = 0$, $\rho = \delta_{N, n_0}$ and therefore $S(T = 0)$ vanishes. %
\end{thebibliography}
\end{document}